\documentstyle[prl,epsf,amssymb,amsbsy,amstext,amsfonts,aps]{revtex}
\begin{document}
\twocolumn
\draft{}
\bibliographystyle{try}

\topmargin 0.1cm


\def\sphn{$^1$}
\def\genova{$^{2}$}
\def\asuaz{$^{3}$}
\def\cmupa{$^{4}$}
\def\cuawdc{$^{5}$}
\def\cnuva{$^{6}$}
\def\cwm{$^{7}$}
\def\edinburgh{$^{8}$}
\def\duke{$^{9}$}
\def\fsu{$^{10}$}
\def\fiu{$^{11}$}
\def\gwudc{$^{12}$}
\def\ipn{$^{13}$}
\def\itep{$^{14}$}
\def\jmuva{$^{15}$}
\def\knukorea{$^{16}$}
\def\frascati{$^{17}$}
\def\mit{$^{18}$}
\def\nsuva{$^{19}$}
\def\ohio{$^{20}$}
\def\oduva{$^{21}$}
\def\rubltx{$^{23}$}
\def\rpi{$^{22}$}
\def\jlab{$^{24}$}
\def\ucla{$^{25}$}
\def\connecticut{$^{26}$}
\def\umma{$^{27}$}
\def\unhdurham{$^{28}$}
\def\uppa{$^{29}$}
\def\urva{$^{30}$}
\def\usc{$^{31}$}
\def\utep{$^{32}$}
\def\uvch{$^{33}$}
\def\vpsu{$^{34}$}
\def\yerevan{$^{35}$}

\title{Photoproduction of $\phi$(1020) Mesons on the Proton at Large Momentum Transfer}

\author{E.~Anciant,\sphn\  
        T.~Auger,\sphn\ 
       G.~Audit,\sphn\ 
         M.~Battaglieri,\genova\  
        J.M.~Laget,\sphn\ 
        C.~Marchand,\sphn\   
        G.S.~Adams,\rpi\ 
        M.J.~Amaryan,\yerevan\ 
        M.~Anghinolfi,\genova\        
        D.~Armstrong,\cwm
% ...
        \\(The CLAS Collaboration)
}


\address{
\sphn CEA Saclay, DAPNIA-SPhN, F91191 Gif-sur-Yvette Cedex, France\\
\genova Istituto Nazionale di Fisica Nucleare, Sezione di Genova
e Dipartimento di Fisica dell'Universita, 16146 Genova, Italy\\
\asuaz Arizona State University, Department of Physics and Astronomy, Tempe, AZ 85287, USA\\
\cmupa Carnegie Mellon University, Department of Physics, Pittsburgh, PA 15213, USA\\
\cuawdc Catholic University of America, Department of Physics, Washington D.C., 20064, USA\\
\cnuva Christopher Newport University, Newport News, VA 23606, USA\\
\cwm College of William and Mary, Department of Physics, Williamsburg, VA 23187, USA\\
\edinburgh Department of Physics and Astronomy, Edinburgh University, Edinburgh EH9 3JZ, United Kingdom\\
\duke Duke University, Physics  Bldg. TUNL, Durham, NC27706, USA\\
\fsu Florida State University, Department of Physics, Tallahassee, FL 32306, USA\\
\fiu Florida International University, Miami, FL 33199, USA\\
\gwudc George Washington University, Department of Physics, Washington D. C., 20052 USA\\
\ipn Institut de Physique Nucleaire d'Orsay, IN2P3, BP 1, 91406 Orsay, France\\
\itep Institute of Theoretical and Experimental Physics, 25 B. Cheremushkinskaya, Moscow, 117259, Russia\\
\jmuva James Madison University, Department of Physics, Harrisonburg, VA 22807, USA\\
\knukorea Kyungpook National University, Department of Physics, Taegu 702-701, South Korea\\
\frascati Istituto Nazionale di Fisica Nucleare, Laboratori Nazionali di Frascati, P.O. 13, 00044 Frascati, Italy\\
\mit M.I.T.-Bates Linear Accelerator, Middleton, MA 01949, USA\\
\nsuva Norfolk State University, Norfolk VA 23504, USA\\
\ohio Ohio University, Department of Physics, Athens, OH 45701, USA\\
\oduva Old Dominion University, Department of Physics, Norfolk VA 23529, USA\\
\rpi Rensselaer Polytechnic Institute, Department of Physics, Troy, NY 12181, USA\\
\rubltx Rice University, Bonner Lab, Box 1892, Houston, TX 77251\\
\jlab Thomas Jefferson National Accelerator Facility, 12000 Jefferson Avenue, Newport News, VA 23606, USA\\
\ucla University of California at Los Angeles, Physics Department, 405 Hilgard Ave., Los Angeles, CA 90095-1547, USA\\
\connecticut University of Connecticut, Physics Department, Storrs, CT 06269, USA\\
\umma University of Massachusetts, Department of Physics, Amherst, MA 01003, USA\\
\unhdurham University of New Hampshire, Department of Physics, Durham, NH 03824, USA\\
\uppa University of Pittsburgh, Department of Physics, Pittsburgh, PA 15260, USA\\
\urva University of Richmond, Department of Physics, Richmond, VA 23173, USA\\
\usc University of South Carolina, Department of Physics, Columbia, SC 29208, USA\\
\utep University of Texas at El Paso, Department of Physics, El Paso, Texas 79968, USA\\
\uvch University of Virginia, Department of Physics, Charlottesville, VA 22903, USA\\
\vpsu Virginia Polytechnic and State University, Department of Physics, Blacksburg, VA 24061, USA\\
\yerevan Yerevan Physics Institute, 375036 Yerevan, Armenia\\
}
\date{\today}
\maketitle

\newpage

\wideabs{
\begin{abstract}
 The cross section for $\phi$ meson photoproduction on the proton has been measured
  for the first time up to a four-momentum transfer $-t = 4$ GeV$^2$, using the CLAS detector
   at the Thomas Jefferson National Accelerator Facility. 
At low four-momentum transfer, the differential cross section is well described 
by  Pomeron exchange. At large four-momentum transfer, above $-t = 1.8$ GeV$^2$,
the data support a model where the Pomeron is resolved into its simplest component, two gluons,
which may couple to any quark in the proton and in the $\phi$.
\end{abstract}

}

\narrowtext

In this paper we report results of the first determination of the cross section
for elastic $\phi$ photoproduction  on the proton, up to $ -t = 4$  GeV$^2$, in a kinematical
domain  where  the  Pomeron  may  be  resolved  into  its  simplest  2-gluon
component. Due to the  dominant  $s\overline{s}$
component of the $\phi$, and to the extent that the  strangeness  component
of the nucleon is small, the exchange of quarks is strongly suppressed.

The scarce existing  experimental  data for 
this reaction~\cite{An,Abbhhm,Be,Ba,Bal} extend only to a momentum transfer of $-t  =  1$ GeV$^2$ and
are well described as a purely diffractive process 
involving the exchange of the Pomeron trajectory in the $t$  channel~\cite{DL87}.
At larger $t$, the small impact  parameter makes it possible for a quark in
the vector meson and a quark in the proton  to  become  close  enough to
exchange  two gluons  which do not have  enough  time to  reinteract  to form a
Pomeron. Such a model of the Pomeron as two non-perturbative gluons~\cite{DL89}
matches the Pomeron model up to $-t = 1$ GeV$^2$, but predicts a different behavior at higher $t$~\cite{La95}.

\begin{figure}[h]
\epsfxsize5. cm
\epsfysize6. cm
\centerline{\epsffile{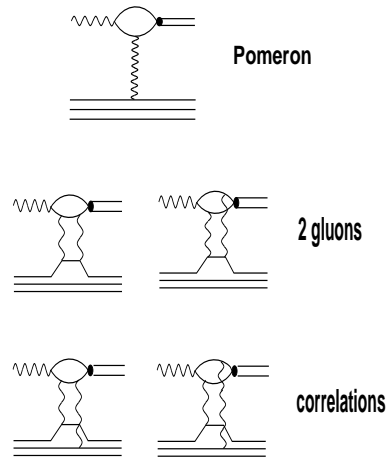}}
\vspace{1cm}
\caption[]{ Diagrams representing the exchange of a Pomeron or of 2 gluons in
the photoproduction of the $\phi$.}
\label{diagrams}
\end{figure}

 Large momentum  transfers
also select  configurations in which the transverse distances between the two quarks
in the vector  meson and the three  quarks in the proton are small.  In that
case,  each   gluon  can   couple  to   different   quarks   of  the   vector
meson~\cite{La95},  as depicted in the middle diagram of  Fig.~\ref{diagrams},
as  well  as to two  different  quarks  of  the  proton~\cite{La98}  (bottom
diagrams in  Fig.~\ref{diagrams}).
So, elastic  $\phi$ photoproduction at large $t$ is a good tool to gain access to the 
quark correlation function in the proton~\cite{La00,Be00,On00}.

  Measurements at such large four-momentum transfers are now possible thanks to the  continuous beam of \mbox{CEBAF} at Jefferson Lab.  This experiment was performed  using  the Hall B tagged photon beam.
  The incident electron beam, with an energy $E_0$ = 4.1 GeV, impinged upon a
  gold radiator of $10^{-4}$ radiation lengths. The tagging system, which gives a  
  photon-energy resolution of 0.1\% E$_0$, is described in Ref.~\cite{SO99}. For this experiment the
  photons were tagged only in the range 3.3-3.9 GeV.
  The target cell, a mylar cylinder  6 cm in diameter and 18 cm long, was filled
  with liquid hydrogen at 20.4 K. 
 
  The photon flux was determined with a pair spectrometer located
  downstream of the target. The efficiency of this pair spectrometer was measured 
  at low intensity ($10^5 \gamma$/s in the entire  bremsstrahlung spectrum) by comparison with a total 
  absorption counter  (a lead-glass detector of 20 radiation lengths).  During data taking at high intensity 
  ($6\times 10^6$ tagged $\gamma$/s), the number of coincidences, true and accidental,
  between the pair spectrometer and the tagger was recorded by scalers. 
  The number of photons lost in the target and  along the beamline was evaluated
  with a GEANT simulation. The correction is of the order of 5\%.
  The systematic uncertainty on the photon flux has been estimated to be 3\%.

  The hadrons were detected in CLAS, the CEBAF Large Acceptance Spectrometer~\cite{Br00}.
  It consists of a six-coil superconducting magnet producing a toroidal field.
  Three sets of drift chambers allow the determination of the momenta of the charged particles with polar
  angles from 10 to 140 degrees. A complete coverage of scintillators allows the discrimination 
  of particles by a time-of-flight technique as described in Ref.~\cite{Sm99}. 
  As the field in the magnet was set to bend the positive particles outwards,
  the $K^-$, from the $\phi \rightarrow K^+K^-$ decay, were identified by the missing mass of the reaction $\gamma p \longrightarrow pK^+(X)$.

\begin{figure}[h]
\epsfxsize8. cm
\epsfysize8. cm
\centerline{\epsffile{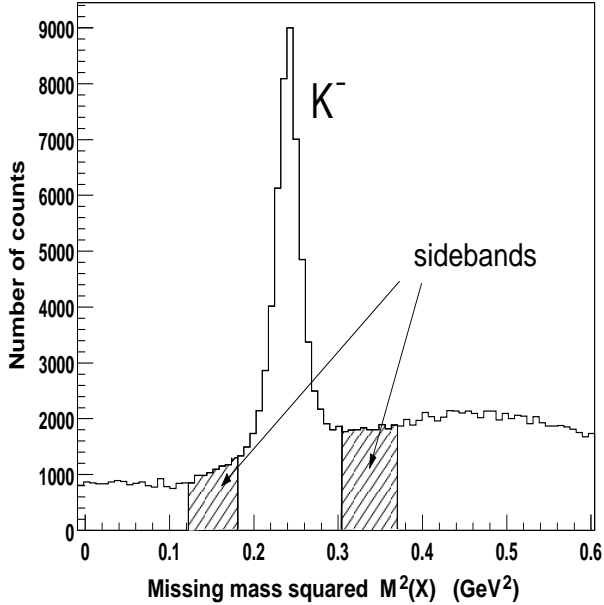}}

\caption[]{ Missing mass squared M$^2$(X) in the reaction $\gamma p \longrightarrow p K^+ (X)$.}
\label{MM}
\end{figure}

\begin{figure}[h]
\epsfxsize8. cm
\epsfysize7.9 cm
\centerline{\epsffile{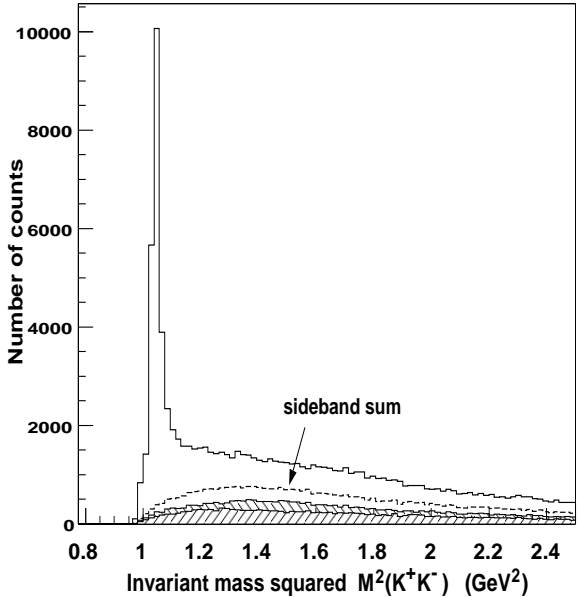}}
\caption[]{The $K^+K^-$ mass spectrum, before the sideband subtraction. Slashes: lower mass sideband contribution. Anti-slashes: higher mass sideband contribution. }
\label{subs}
\end{figure}

  In Fig.~\ref{MM}, a well-identified $K^-$ peak can be seen
  above a background which corresponds to a combination of misidentified particles, the contribution 
  of multi-particle channels and accidentals between CLAS and the tagger.
  The background is eliminated by subtracting the counts in the sidebands, indicated in the figure, from the 
  main peak, in each bin in $t$ (determined by the four-momentum of the detected proton). The contribution of the sidebands to the $K^+K^-$ mass spectrum is shown in Fig.~\ref{subs}. Note  that it is very small under the $\phi$ peak.

\begin{figure}[h]
\epsfxsize8. cm
\epsfysize8. cm
\centerline{\epsffile{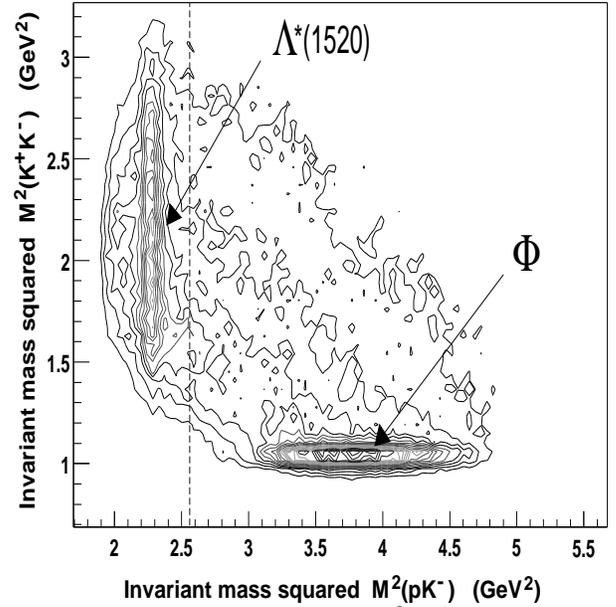}}
\caption[]{ Invariant mass squared M$^2$($K^+$$K^-$) as function of M$^2$(p$K^-$).}
\label{dalitz}
\end{figure}

In the Dalitz plot (Fig.~\ref{dalitz})  of invariant masses squared M$^2$($K^+$$K^-$) versus M$^2$($pK^-$), 
  two resonant contributions to 
  the $pK^+$$K^-$ channel can be clearly  seen, namely the $p\phi$ and
   the $\Lambda^{*}(1520)K^+$ channels.
  A cut at M$^2$($pK^-$) $> 2.56$ GeV$^2$ further suppresses the contribution of the $\Lambda^*$ production to the $K^+K^-$ mass spectrum.

\begin{figure}[h]
\epsfxsize=8. cm
\centerline{\epsffile{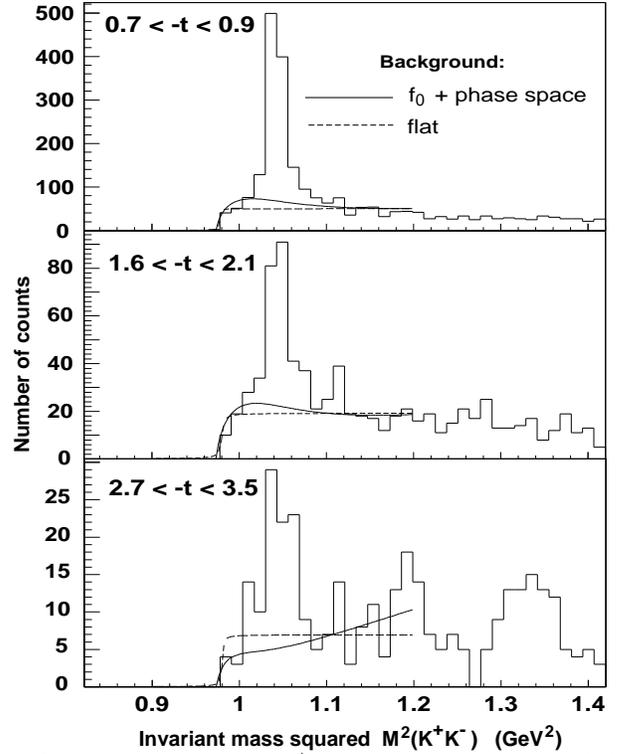}}
\caption[]{Invariant mass $K^+K^-$ for selected values of the four-momentum transfer $t$ (GeV$^2$), after sideband subtraction. The curves show the  continuum obtained from the fits discussed in the text.}
\label{tdis}
\end{figure}  

  The resulting mass spectra are shown in Fig.~\ref{tdis} for selected bins in $t$.
  The peak of the $\phi$(1020) clearly shows up  over a  $K^+$$K^-$ continuum contribution 
  which must be subtracted. The $\phi$ events are selected by the cut  $ 1.0 < M^{2}(K^+K^-) < 1.1$ GeV$^2$.
    The CLAS acceptance in the forward direction limits the data set to values of $-t$ larger than 
  $0.4$ GeV$^2$. This experiment extends the measured range up to $-t= 4$ GeV$^2$.  
  
The detector efficiency depends on four variables: $E_\gamma$, $t$, $\theta^{cm}_{K^+}$ and $\phi^{cm}_{K^+}$ (the decay angles of the $K^+$ in the c.m. of the $\phi$). A GEANT simulation program, which takes into account the entire CLAS setup, was used to calculate the detector efficiency,  taking into account in an iterative way the experimentally observed variation of the cross section as a function of these variables. No variations of the cross-section against $E_\gamma$ and $\phi^{cm}_{K^+}$ were observed. This efficiency varies from 0.15 to 0.25. The accuracy of the simulation has been evaluated to be 5\%  from a comparison between the real data  and the Monte Carlo simulation~\cite{Aug}  for the  channel $\gamma p \longrightarrow p \pi^+ \pi^-$,  where the statistics are very high.

The continuum background  has been subtracted assuming an isotropic distribution in $\theta_{K^+}^{cm}$
  and two hypotheses for its variation against the mass $M(K^{+}K^{-}$):
  i)~a flat contribution, and 
  ii)~a phase space distribution plus a contribution of the $f_{0}(980) $ decaying into two kaons
       (the mass of the $f_{0}$ is below the two-kaon threshold but because of its $\sim$~60 MeV width,
        the tail of the Breit-Wigner can contribute). 
Its contribution was determined by fitting the $K^+K^-$ mass spectrum  (up to $M^{2}(K^+K^-) = 1.2$ GeV$^2$ in each bin in $t$)  with two components: the background itself and a Breit-Wigner describing the $\phi$ meson peak.

The results for the cross section are the average between the two values obtained according to
these two background hypotheses, with the difference being taken 
as an estimate of the systematic uncertainty due to the subtraction of the $K^{+}K^{-}$  continuum 
production. The data are integrated over the full tagging energy range ($3.3$ GeV $< E_{\gamma} < 3.9 $ GeV).
  
\begin{figure}[h]
\epsfxsize=8. cm
\centerline{\epsffile{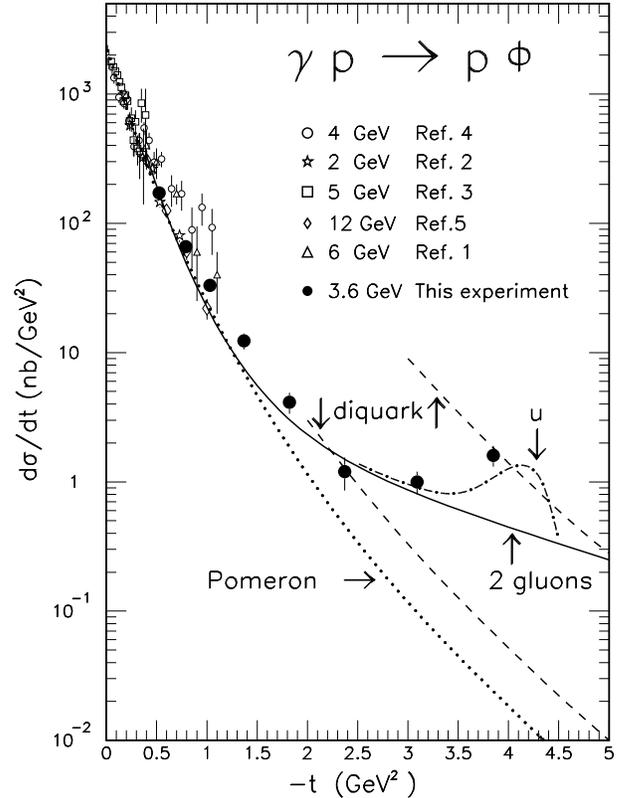}}
\caption[]{The differential $\phi$ photoproduction cross-section versus 
         the four-momentum transfer $t$ (see text for the explanation of the curves).
         The error bars displayed are the quadratic sum of statistical and systematic uncertainties
         which include 3\% for normalization, 5\% for acceptance and 5-15\% for background subtraction.}
\label{sig_tot}
\end{figure}

 The cross sections $d\sigma/dt$ versus $t$ for the $\phi$ photoproduction are presented in Fig.~\ref{sig_tot}, for eight bins in $t$.
  For values of $-t$ around $1$~GeV$^2$, our data are in good agreement 
  with the most precise  published data. The dotted  curve  corresponds  to Pomeron  exchange~\cite{La00}.  The solid curve  corresponds  to the exchange of two  non-perturbatively dressed gluons~\cite{La98,La00} 
that may couple  to any quark  in the $\phi$ meson and in the  proton. It includes quark  correlations  in the proton, assuming the simplest form of its wave  function~\cite{Cu94}:
three valence quarks equally sharing the proton  longitudinal  momentum. The parameters in this model are
fixed by the analysis of other independent channels. It also reproduces  the data recently recorded at HERA~\cite{Hera} up to $-t = 1$~GeV$^2$ (see Ref.~\cite{La00}).
 
The solid curve gives a good  description  of the  experiment over the entire  range of $t$ except for the last point at $-t=3.9$ GeV$^{2}$. Here, one approaches the kinematical limit and $u$-channel nucleon
exchange may contribute~\cite{La00}.
Performing the experiment at higher average energy (4.5 GeV) would push the 
$u$-channel contribution to higher values of $|t|$ (6 GeV$^2$) and leave a wider window to study two-gluon exchange mechanisms. 

 The dot-dashed curve includes the $u$-channel contribution with
 the choice $g_{\phi NN}=3$  for the $\phi NN$ coupling (the addition of the $u$-channel amplitude to the
 dominant $t$-channel  amplitude does not lead to double counting, because the former relies on quark exchange and the latter relies on gluon exchange). 
This value comes from  the analysis 
of nucleon electromagnetic form factors~\cite{Jaffe} as well as  nucleon-nucleon and
hyperon-nucleon scattering~\cite{Nag}. It is higher than the value $g_{\phi NN} = 1$ predicted from SU(3) 
mass splitting or  $\omega-\phi$ mixing~\cite{Jain}, thus confirming evidence for additional OZI-evading processes at the $\phi NN $ vertex.

The   predictions  of two other models are also presented in Fig.~\ref{sig_tot}. Both treat
the gluon exchange in perturbative QCD (this leads to the characteristic $t^{-7}$ behavior of a hard scattering) and use a diquark model to take into account
 quark correlations in the proton (this fixes the magnitude of the cross section). Berger   and
Schweiger~\cite{Be00} (upper dashed curve) use a wave function which leads to a good accounting of Compton scattering and nucleon form factors,  while  Carimalo {\it et  al.}~\cite{On00} (lower dashed curve) use a wave function which fits the cross section of the $\gamma \gamma \rightarrow p\overline{p}$ reaction. 
Above $-t=2$ GeV$^2$ our data rule out the $t$  dependence of these
diquark models,  demonstrating that the asymptotic regime is not yet reached.
Recently, a new anomalous Regge trajectory associated with the $f_{1}$(1285) meson has been proposed~\cite{Ko99}.
It reproduces the HERA~\cite{Hera} data ($-t < 1$ GeV$^2$), but its momentum dependence is too steep to reproduce our high $t$ data.

\begin{figure}[h]
\epsfxsize=7. cm
\centerline{\epsffile{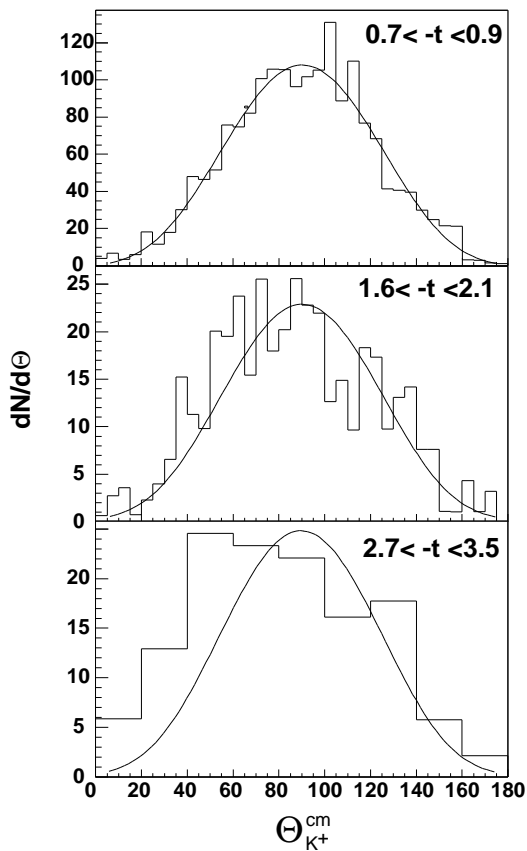}}
\caption[]{Angular distributions (corrected for acceptance) $dN/d\theta$ of the $K^+$, in the helicity frame, are compared to the prediction of SCHC.}
\label{thcm}
\end{figure}
  
Figure~\ref{thcm} shows the decay angular distributions of the $\phi$ in the helicity frame~\cite{Sc}
for selected bins in $t$.  The sideband contributions have been subtracted, but not  the $K^+K^-$ continuum contribution. Up to $-t = 2.1$ GeV$^2$
they follow a $\sin^{2}\theta d(\cos\theta)$ dependence, in agreement with $s$-Channel Helicity Conservation (SCHC): a real photon produces a $\phi$ meson with only transverse components.
 Above $-t=2.7$ GeV$^2$, there is a  violation of SCHC, likely to be 
associated with the $u$-channel exchange and the interference between the $\phi$ and the S-wave $K^+K^-$
photoproduction amplitudes.

  In conclusion, elastic photoproduction of $\phi$ mesons from the proton was measured for
  the first time up to $-t=4$ GeV$^2$.   Below $-t\approx 1$ GeV$^2$, they cannot distinguish between the Pomeron exchange  and the 2-gluon exchange models which both agree with the existing data.  At high $t$, the predictions of these models differ by more than an order of magnitude.  Above $-t\approx 1.8$ GeV$^2$, our  data rule out the diffractive Pomeron and strongly favor its  2-gluon realization.  This opens a window to the study of the quark correlation function in the proton.

We would like to acknowledge the outstanding efforts of the staff of the Accelerator and the Physics Divisions at JLab that made this experiment possible. This work was supported in part by the French Commissariat \`a l'Energie Atomique, the Italian Istituto Nazionale di Fisica Nucleare, the U.S. Department of Energy and National Science Foundation, and the Korea Science and Engineering Foundation.

\end{document}